%% file: 0_paper_acm.tex
\documentclass[sigconf]{acmart}
\usepackage[utf8]{inputenc}
\usepackage[ngerman,english]{babel}
\usepackage{todonotes}
\usepackage{booktabs,graphicx}
\usepackage{float}
\usepackage{enumitem}
\usepackage{array}
\AtBeginDocument{%
  \providecommand\BibTeX{{%
    \normalfont B\kern-0.5em{\scshape i\kern-0.25em b}\kern-0.8em\TeX}}}

\copyrightyear{2020}
\acmYear{2020}
\setcopyright{acmlicensed}\acmConference[JCDL '20]{Proceedings of the ACM/IEEE Joint Conference on Digital Libraries in 2020}{August 1--5, 2020}{Virtual Event, China}
\acmBooktitle{Proceedings of the ACM/IEEE Joint Conference on Digital Libraries in 2020 (JCDL '20), August 1--5, 2020, Virtual Event, China}
\acmPrice{15.00}
\acmDOI{10.1145/3383583.3398534}
\acmISBN{978-1-4503-7585-6/20/06}

\begin{document}

\title{HybridCite: A Hybrid Model for Context-Aware Citation Recommendation}

\author{Michael Färber}
\orcid{0000-0001-5458-8645}
\affiliation{%
  \institution{Karlsruhe Institute of Technology (KIT)}
  \city{Karlsruhe}
  \postcode{76133}
    \country{Germany}
}
\email{michael.faerber@kit.edu}

\author{Ashwath Sampath}
\affiliation{%
  \institution{University of Freiburg}
  \streetaddress{Georges-Koehler-Allee 051}
  \city{Freiburg}
  \postcode{79110}
    \country{Germany}
}
\email{ashwath92@gmail.com}

\newcolumntype{R}[1]{>{\raggedleft\arraybackslash}p{#1}}

\begin{abstract}
Citation recommendation systems aim to recommend citations for either a complete paper or a small portion of text called a citation context. The process of recommending citations for citation contexts is called local citation recommendation and is the focus of this paper. Firstly, we develop citation recommendation approaches based on embeddings, topic modeling, and information retrieval techniques. We combine, for the first time to the best of our knowledge, the best-performing algorithms into a semi-genetic hybrid recommender system for citation recommendation. We evaluate the single approaches and the hybrid approach offline based on several data sets, such as the Microsoft Academic Graph (MAG) and the MAG in combination with arXiv and ACL. 
We further conduct a user study for evaluating our approaches online. Our evaluation results show that a hybrid model containing embedding and information retrieval-based components outperforms its individual components and further algorithms by a large margin. 
\end{abstract}


\maketitle

\input{1_introduction.tex}
\input{2_relatedwork.tex}
\input{3_approach.tex}
\input{4_evaluation.tex}
\input{5_conclusion.tex}

\bibliographystyle{ACM-Reference-Format}
\bibliography{bibliography}

\end{document}

%% file: 1_introduction.tex
\section{Introduction}
\label{sec:introduction}

Citations are the lifeblood of academic research papers. They provide a measure of trustworthiness and can be used to either back up the authors’ claims using earlier research, to improve upon existing methods, or even to criticize research done previously. 

However, as the number of new scientific publications has been on a steep upward curve in recent years (detailed statistics in \cite{stm2008}), the researcher's task of finding a suitable paper to refer to and cite has become much more challenging than ever. As a result, an increased amount of research is now being put into citation recommendation \cite{CiteRecSurvey} -- the process of finding and recommending prior work based on a passage in the text. This text passage, usually called the \textit{citation context}, can be of different lengths, ranging from a phrase or sentence up to a whole document.

Works on global recommendations, that is, citations for the entire paper, have been carried out by ~\cite{McNeeACGLRKR02}, ~\cite{StrohmanCJ07}, and \cite{NallapatiAXC08}, and more recently by ~\cite{Bhagavatula2018}, ~\cite{YangZCDMGD18}, and \cite{CaiHY18}. This paper, however, focuses on \textit{local citation recommendation}, in which a relatively small citation context of 1-3 sentences or 50-100 words is used as input for the recommendation. This type of fine-grained recommendation, sometimes also referred to as \textit{context-aware citation recommendation} in contemporary research papers, was first explored in \cite{He2010} and \cite{He2011}. 

Previous works also include personalized approaches, such as \cite{Ebesu2017} and \cite{YangZCDMGD18}, which use author and venue metadata as input and generally obtain better scores in evaluations. However, as explained in \cite{Bhagavatula2018}, this is due to the fact that metrics usually favor predictions of obvious citations resulting in better scores, for example citations by the same author. Therefore, this paper will not consider such personalized approaches.

In this paper, we, first of all, adapt an existing deep-learning-based embedding method to citation recommendation (Hyperdoc2vec by Han et al. \cite{ShiSZZH18}). In addition, we develop several baselines based on topic modeling and classical information retrieval, such as a BM25-based approach, an approach based on Latent Dirichlet Allocation (LDA) \cite{BleiNJ03}, and one based on paragraph vectors \cite{LeM14}. 
Secondly and more importantly, we combine the best-performing recommendation methods of the previous step into a weighted hybrid recommender system for the task of citation recommendation. 

While existing approaches to citation recommendation can be broken down into two steps (see, e.g.,~\cite{Bhagavatula2018}), to the best of our knowledge, no approach has been truly hybrid in nature, i.e., combines results from two different recommendation algorithms. Even though hybrid recommendation approaches have been proposed in other fields, such as for paper recommendation~\cite{Kanakia2019WWW}, the task differs considerably from local citation recommendation, as paper recommendation does not consider citation contexts, leading to different system setups and evaluation setups~\cite{He2010}.

To conduct our experiments, we use, for the first time in the area of (local) citation recommendation, the rich Microsoft Academic Graph (MAG) as one of our data sources. We also prepare two auxiliary data sets based on the MAG, with restrictions made on the language and discipline (English and computer science, respectively) -- the arXiv data set~\cite{SaierF19} and the ACL-ARC data set~\cite{BirdDDGJKLPRT08}. These are mapped back to the MAG and made publicly available. Overall, we create five large-scale evaluation data sets. 

We then evaluate all of our proposed baselines and approaches on the different data sets. 

\newpage
Overall, we make the following contributions:

\begin{itemize}
    \item We present a hybrid approach to citation recommendation that combines individual approaches to citation recommendation stochastically.\footnote{The source code is available online at \url{https://github.com/ashwath92/HybridCite}.}
    \item We prepare two conceptually different data sets (based on citing and cited papers) that will be used in conjunction in an advanced hybrid recommender system. Our advanced hybrid recommender system thus combines several algorithms as well as several data sets.\footnote{See \url{https://github.com/ashwath92/HybridCite}.}
    \item We perform an extensive offline evaluation of the developed approaches based on five data sets. Among other data sets, we prepare the MAG (with over 1.6M computer science papers) to this end and provide it online. 
    In addition, we perform a user study. In all evaluations, we can show the superiority of the presented hybrid recommender over its individual components. 
\end{itemize}

The rest of our paper is structured as follows: 
After giving an overview of related work on citation recommendation in Section~\ref{sec:relatedwork}, we present our new approaches to citation recommendation in Section~\ref{sec:approach}. Section~\ref{sec:evaluation} outlines the evaluation setup and the evaluation results. We conclude in Section~\ref{sec:conclusion} with a conclusion and an outlook.

%% file: 2_relatedwork.tex
\section{Related Work}
\label{sec:relatedwork}

McNee~\cite{McNeeACGLRKR02} in 2002 and Strohman et al.~\cite{StrohmanCJ07} in 2007 published the first global citation recommendation papers. Since then, various papers on both global and local citation recommendation have been published.

\textbf{Local Citation Recommendation}
The term \textit{context-aware recommendation} was introduced by He et al.~\cite{He2010} in 2010, the first paper which covered local recommendation.
The authors expanded their model in \cite{He2011}. 
Huang et al.~\cite{HuangKCMGR12} built upon the idea by translating specific keywords in the contexts (source language) into cited documents (target language), thereby creating a de-facto machine translation system for citation recommendation. 

Huang et al.’s paper~\cite{Huang2015} was one of the follow-up papers to Huang et al.~\cite{HuangKCMGR12}. Here, they continued their work on translation models, but added in distributed word representations of the words and cited documents in the citation contexts (see~\cite{MikolovSCCD13}). 

\textbf{Embedding-based Approaches}
Tang et al.~\cite{TangWZ14} introduced embedding-based approaches to the field of citation recommendation by using TF-IDF vectors to construct cross-language embeddings for local citation recommendation. Jiang et al.'s two papers \cite{JiangLL18,JiangYGLL18} also used embeddings in the context of cross-language global citation recommendation. Similar works were carried out by Cai et al.~\cite{CaiHY18} and Zhang et al.~\cite{ZhangYCD18} in 2018.

A recent paper on embedding-based neural networks by Han et al.~\cite{ShiSZZH18} emphasized content awareness, context awareness, newcomer friendliness, and context intent awareness. Due to these characteristics, we have adapted their approach in this paper.

\textbf{Topic Modelling and Information Retrieval}
Topic modelling and, in particular, Latent Dirichlet Allocation (LDA)~\cite{BleiNJ03} have been used in multiple citation recommendation papers \cite{KatariaMB10, DBLP:Jiang13, NallapatiAXC08, LiuYGSG14}. In this paper, LDA is used as a baseline.

Information retrieval techniques such as TF-IDF-based text comparison or BM25 have been studied previously. Duma et al.'s two papers \cite{DumaKLRC16, DumaLCRK16} treated citation recommendation as an information retrieval task, whereas Ebesu et al.~\cite{Ebesu2017} used BM25 as a simple baseline. We use BM25 in our hybrid recommendation system.

\textbf{Hybrid recommender systems}
Hybrid recommender systems take predictions from two or more disparate recommender systems and combine them in some way. 

Burke \cite{Burke2002, Burke2007} provided an excellent introduction and survey about hybrid recommender systems. In the citation recommendation context, Hsiao et al.~\cite{HsiaoCD15} used a hybrid recommendation system that combines results from two disparate systems. The authors looked for recommendations from one algorithm and only selected the recommendations of the other algorithm as a last resort. In our paper, however, we create a stochastic (semi-genetic) hybrid recommender system which combines results from multiple sources. 

A two-step process (candidate generation, ranking) to citation recommendation is explored in several papers, including Zarrin\-kalam~\cite{Zarrinkalam2013}, Bhagavatula et al.~\cite{Bhagavatula2018}, and McNee~\cite{McNeeACGLRKR02}, but these are not hybrid systems per se -- as they merely use two different algorithms for candidate generation and ranking. Kanakia et al.~\cite{Kanakia2019WWW} presented a hybrid \textit{paper} recommender system in which they combined a co-citation-based algorithm and a content-based algorithm. Finally, Rokach et al.~\cite{Rokach2013} present a hybrid citation recommendation system based on multiple machine learning algorithms whose results are combined by simple averaging. In this paper, we use a semi-genetic algorithm to stochastically combine results from different categories of algorithms while using multiple data sets (concerning citing and cited papers). Due to this incompatibility, a direct comparison to these approaches is not possible. 

%% file: 3_approach.tex
\section{Approach}
\label{sec:approach}

In this section, we outline the single approaches used as components as well as the hybrid recommendation algorithm itself.

\subsection{Single Approaches}
\subsubsection{BM25}

Okapi BM25, a bag-of-words algorithm, has been used in citation recommendation approaches both as a pre-filter and as a simple baseline \cite{Ebesu2017}.
BM25 ranks returned documents based on the query terms appearing in each document. However, the specific position of the query terms makes no difference to the ranking algorithm. 

\begin{figure*}
 \includegraphics[width=16cm]{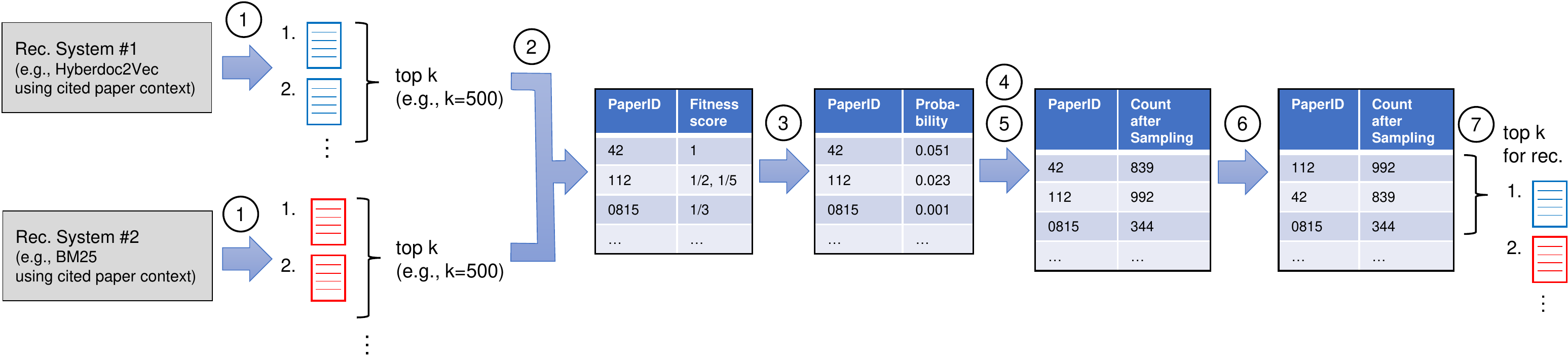}
  \caption{Overview of the hybrid semi-genetic hybridisation algorithm for citation recommendation. The encircled numbers indicate the main steps of the algorithm as outlined in Sec.~\ref{sec:hybrid-rec-approach}.}
  \label{fig:hybridv2flowchart}
\end{figure*}

\subsubsection{Latent Dirichlet Allocation (LDA)}

The main idea behind using LDA~\cite{BleiNJ03} for citation recommendation is to recommend the same citations for the same topics (given by the citation contexts).
The topics generated via LDA from the citation context are compared to all topics using cosine similarity. The similarities are sorted in descending order, and the top $n$ similar papers are recommended.

\subsubsection{Doc2Vec} 
The literature has many deep learning approaches for citation recommendation, such as RNNs~\cite{Kobayashi2018, YangZCDMGD18} and CNNs~\cite{Ebesu2017, YinL17}. 
In contrast, embedding-based approaches need considerably less time and resources to train while still achieving reasonable results in recommendation tasks~\cite{ShiSZZH18}. 
One embedding approach is doc2vec~\cite{LeM14}, which is an extension of word2vec. 
We train our own embeddings using the respective data sets and generate doc2vec vectors (paragraph vectors) as implicit knowledge representations for the candidate papers and the citation contexts. We can then use the cosine similarity for finding the nearest doc2vec paper embedding for a given doc2vec citation context embedding.

\subsubsection{Paper2Vec} Paper2Vec~\cite{GangulyP17} is another embedding approach to produce document vectors (paragraph vectors) enriched by random walks. Embeddings are trained by combining the papers' textual information and citation information in two distinct steps. The underlying idea is that papers with semantically similar contents are placed closely together. This is then used to obtain a ranking of similar papers for recommendation. 
Note that Paper2Vec is used in the evaluation as a baseline but is not further described in this section, since the approach is described sufficiently in \cite{GangulyP17}.

\subsubsection{HyperDoc2Vec}

The Hyperdoc2Vec approach~\cite{ShiSZZH18} is a general recommendation approach for hyper-documents. It can thus be applied to citation recommendation. The algorithm produces two vectors for each paper: an IN vector and an OUT vector. The idea of using dual word embeddings originates from Nalisnick et al.~\cite{NalisnickMCC16}, who claimed that two vectors work better than one for a variety of tasks. 
Specifically, for a paper $P$, the IN document vector ($d_I$) represents $P$ playing the role of a source (citing) document. The OUT document vector ($d_O$) represents $P$ playing the role of a target (cited) document. Essentially, this means that the algorithm is both content-aware and context-aware. Both the content of $P$ (which is, in our case, either the paper's full text or, if the full text is not available, a pseudo full text consisting of a title, abstract, and citation contexts where it cites other papers) and the contexts of the papers which cite $P$ play a role in its embeddings. 
The trained paper embeddings can then be compared via a similarity function (e.g., cosine similarity) to find the most suitable papers for a given citation context.

\subsection{Semi-Genetic Hybrid Recommender for Citation Recommendation}
\label{sec:hybrid-rec-approach}

In this subsection, we describe an approach which uses the concept of random draws with replacement to probabilistically integrate results from different algorithms.  

Recommendations from different recommender systems can be combined in several ways. 
One possibility is to use a stochastic weighted hybrid algorithm as described by Mueller~\cite{Mueller17}. The algorithm is called ``semi-genetic'' because the cross-over and mutation steps of genetic algorithms are skipped. In addition, the semi-genetic algorithm has only one iteration, unlike most genetic algorithms. In the following, we adapt this hybrid algorithm for citation recommendation due to its added layer of stochasticity compared to a simple weighted approach.

\textbf{Hybrid} The structure of our hybrid algorithm is illustrated in Figure~\ref{fig:hybridv2flowchart} using 1 data set and 2 recommender systems as components. Thus, we call it \textit{Hybrid12}, or simply \textit{Hybrid}. 
Note that the algorithm works for any number of single recommendation approaches ($m$ in the following) and data sets. 
The workflow is as follows: 

\begin{enumerate}
\item \textit{Initialize population}: select a set of items from all possible ``chromosomes,'' i.e. from the recommendation lists of both component algorithms. In this step, we obtain the top $k$ recommendations (e.g., $k=500$) from each algorithm and concatenate them.

\item \textit{Evaluate fitness scores}: the reciprocal rank is assigned as the fitness score for the recommendations from each of the algorithms. If a paper has been recommended by both algorithms, it will carry two associated scores.

\item  \textit{Convert the scores into probabilities}: this is done by dividing each score by the sum of all the scores. In case of multiple scores, the scores are first summed up. 

\item  \textit{Selection}: we randomly draw $n$ (e.g., $n$ = 1 million) samples with replacement from the array of $l$ ($l=mk$; e.g., $l=1000$) recommendations, based on the probabilities calculated in step 3.

\item \textit{Count} the number of times each of the papers is drawn. 
 
\item \textit{Sort} the recommendations in descending order of frequencies.

\item \textit{Solution set}: return the top $k$ recommendation from the sorted array of recommendations.
\end{enumerate}

\textbf{Hybrid23} Because the papers' full texts are often not available, we introduced so-called \textit{pseudo full texts} of the papers, composed of the papers' titles, abstracts, and citation contexts which are often available (see Section~\ref{sec:evaluation-datasets}). 
However, one can imagine that the citation contexts describe aspects of the \textit{cited} papers they link to rather than the papers in which they appear (called \textit{citing papers} here). For instance, in Figure~\ref{fig:magcited}, the citation contexts in papers a, b, and c describe paper i better than the citation contexts in paper i itself. This assumption is also stated by Huang et al. \cite{HuangKCMGR12,Huang2015}. 

Thus, we now consider the scenario in which -- besides the title and the abstract -- the citation contexts of the \textit{citing papers} (which link to the to-be-modeled-paper) are used (i.e., citation contexts of papers a, b, and c in Figure \ref{fig:magcited}).
For \textit{Hybrid}, preliminary results indicated that the combination of BM25 and the Hyperdoc2vec with OUT document vectors (hd2vOUT) as components of the hybrid recommender system leads to the best results (see Section~\ref{sec:evaluation}). 
We also choose these approaches for the specific set-up in which the citation contexts of the citing paper and cited papers are available. It makes sense to apply hd2vOUT directly on the papers to be modelled (cited papers) but not on the citing papers. Thus, we end up with the following three recommender systems as components of our advanced hybrid approach:

\begin{enumerate}
    \item hd2vOUT trained on a pseudo full text data set containing titles, abstracts, and citation contexts from the \textit{cited} papers; 
    \item BM25 trained on a pseudo full text data set containing titles, abstracts, and citation contexts from the \textit{cited} papers; 
    \item BM25 trained on a pseudo full text data set containing titles, abstracts, and citation contexts from the \textit{citing} papers.
\end{enumerate}

Since this adapted hybrid citation recommendation system is based on two data sets (containing citation contexts from citing papers and citation contexts from the to-be-modeled (i.e., cited) papers) and contains three components, we call it \textit{Hybrid23} in the remaining paper.

%% file: 4_evaluation.tex
\section{Evaluation}
\label{sec:evaluation}

Citation recommender systems are arguably more difficult to evaluate than other types of recommender systems due to the context-awareness and high number of candidate papers (e.g., 1+M). The evaluation depends heavily on the ground truths which are chosen. In the case of an \textit{offline} evaluation, the ground truths are the papers cited by the original authors in the test citation contexts. Thus, the task is to re-predict the citations. However, this kind of evaluation is arguably subjective. There is no reason to believe that all the test contexts' authors have cited the correct papers. For this reason, we conduct an \textit{online} evaluation (user study) in conjunction with an offline evaluation (see Section~\ref{sec:online-evaluation}).

\subsection{Evaluation Data Sets}
\label{sec:evaluation-datasets}

\begin{figure}[tb]
    \centering
    \includegraphics[keepaspectratio, width=\linewidth]{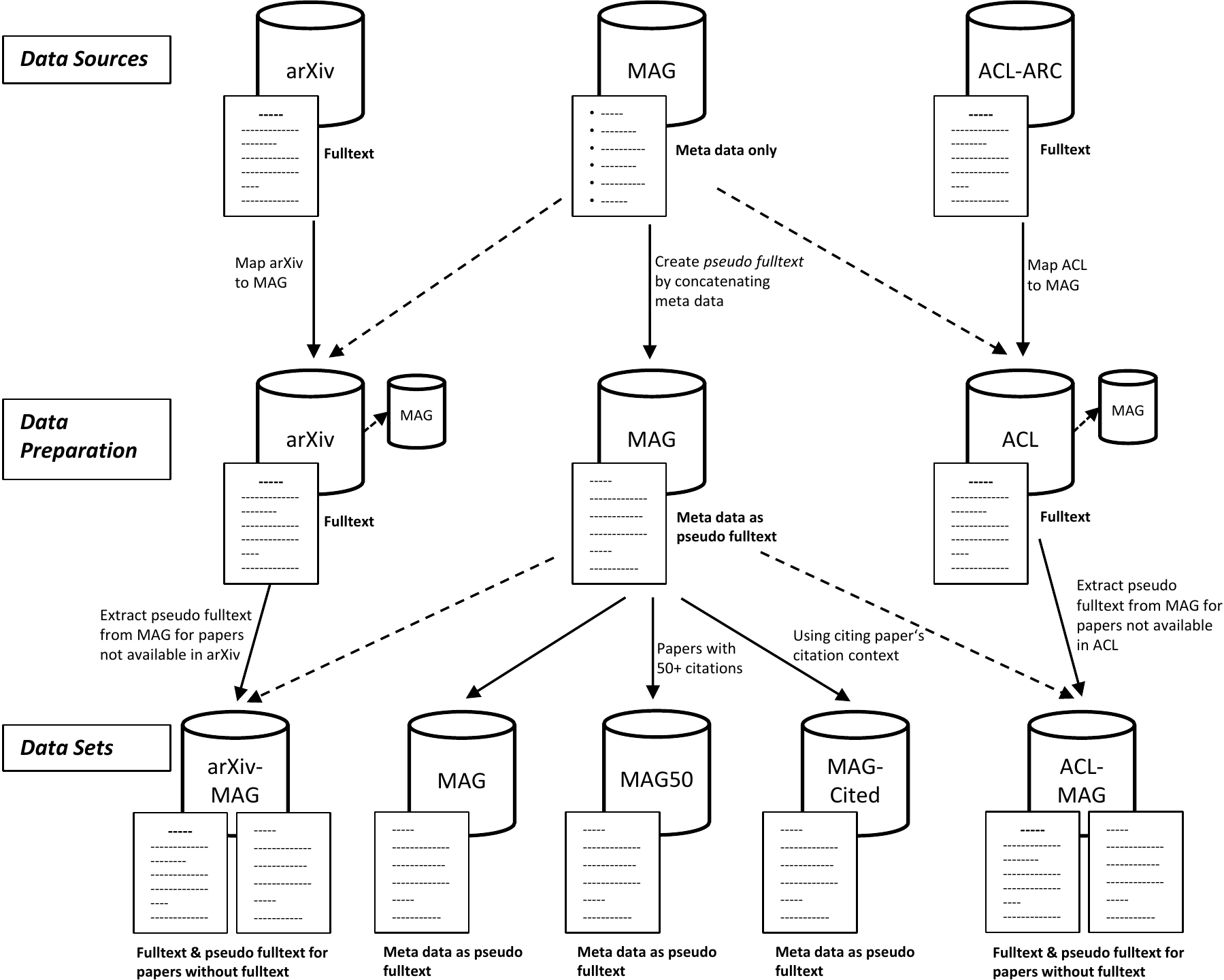}
    \caption{Data set creation process.}
    \label{fig:datasetcreation}
\end{figure}

The amount and quality of real-world data used for training a machine learning model is arguably more important than the quality of the actual algorithms used. Thus, the preparation of the input data is a critical step in the whole process. In the following subsections, we describe our data sources, our data preparation steps, as well as our final data sets used. The entire process is depicted in Figure~\ref{fig:datasetcreation}.

\subsubsection{Data Source}
The following three data sources are used in this paper.\footnote{Note that other data sets containing papers' metadata and citation contexts exist (e.g., RefSeer~\cite{Huang2014}). However, we found the quantity and quality of the papers' metadata and citation contexts in MAG to be much higher. Furthermore, in case of RefSeer, not much data is available for the cited papers which makes it unfit for our approaches.} 

\textbf{Microsoft Academic Graph (MAG)}
The MAG~\cite{Sinha2015}, a relatively new data set, contains metadata (title, abstract, sometimes citation contexts, etc.) from 220 million papers (as of 22 May 2019). Note that the MAG provides no full texts. Papers in the data set range all the way from the years 1800 to 2019.

\textbf{arXiv}
Although arXiv\footnote{\url{https://arxiv.org/}} does not quite match the breadth of coverage in the MAG, it still contains a substantial number of papers (overall over 1.5 million submissions as of 22 May 2019). The papers are from mainly five disciplines -- mathematics, computer science, physics, biology, and economics. 
While citation contexts are not directly available, the full text in PDF as well as \LaTeX{} files can be obtained for almost all arXiv papers.
We use the plain texts extracted from \TeX{} source files as provided by Saier and Färber~\cite{SaierF19}. In this way, we obtain full and clean text data of all arXiv papers.

\textbf{ACL Anthology (ACL-ARC)}
The ACL Anthology Reference corpus is a comparatively tiny data set which covers only computational linguistics and natural language processing. It contains close to 11,000 papers and stretches from 1965 to 2006. The ACL data we use for this paper has been provided in a preprocessed format by Färber et al.~\cite{Faerber2018}, containing sentences, reference information, and publications' metadata.

\subsubsection{Data Set Preparation}
\label{sec:data-set-creation}

The first step was to restrict the data sets to only include papers from the field of computer science. Advantages to using only one discipline include likely better recommendations and reduced training time for embedding-based algorithms while missing out on cross-domain recommendations. 

The MAG data set does not contain full text but rather citation contexts and plenty of metadata in a good quality which needs to be prepared for our content-based algorithms. Therefore, a mechanism is needed to combine the contexts and metadata into a form usable by our algorithms. Thus, we retrieved and concatenated the title, abstract, and citation contexts for each paper to form text which acts as a substitute and which we call the \textit{pseudo full text}. 

In case of the non-MAG data sets, the papers' full text was available. Nevertheless, we mapped the papers to the MAG to enrich their metadata with additional citation information as far as it was missing. 
Title, abstract, and the citation contexts are retrieved and concatenated to form pseudo full text. The reason we go through this process is that one of the used embedding algorithms (Hyperdoc2vec) requires representations for every citing and cited paper, but we cannot ensure that the full texts of all cited and citing documents will be available because they might not be covered by arXiv/ACL-ARC. 

\begin{table}[tb]
    \centering
        \caption{Details about training and test sets.}
    \label{tab:trainingteststats}
    \begin{small}
    \begin{tabular}{lR{1.3cm}R{1.3cm}R{1.2cm}R{1.2cm}}
    \toprule
       \textbf{Data Set}  & \textbf{Training Years} & \textbf{Test Years} & \textbf{\#Training Papers} & \textbf{\#Test Contexts} \\
       \midrule
       ACL-MAG  & 1965-2005 & 2006 & 11,217 & 2,775 \\
       arXiv-MAG  & 1991-2016 & 2017 & 376,218 & 286,272 \\
       MAG  & 1800-2017 & 2018-2019 & 1,620,841 & 168,700 \\
       MAG50  & 1800-2017 & 2018-2019 & 126,666 & 107,781 \\
       MAG-Cited  & 1800-2017 & 2018-2019 & 1,478,924 & 190,991 \\
       \bottomrule
    \end{tabular}
    \end{small}
\end{table}

\subsubsection{Data Sets}
\label{sec:datasets}

Table \ref{tab:trainingteststats} gives an overview of the train/test splits of the used evaluation data sets. 

\textbf{MAG}
To prepare the MAG \textbf{training set}, the dump files are loaded into several PostgreSQL tables. From there, the pseudo full text is prepared for each paper.
The \textbf{test set} uses citation contexts from computer science papers in English which were published in 2018 and 2019.

\textbf{MAG-50}
Here, the data preparation is very similar to the regular MAG data set, the difference being that only papers which have been cited at least 50 times are included in the training set. This means that papers with fewer citations, such as newly published papers, will not get recommended. This data set is used for comparison and analysis purposes. 

\begin{figure}[tb]
    \centering
    \includegraphics[keepaspectratio, width=0.85\linewidth]{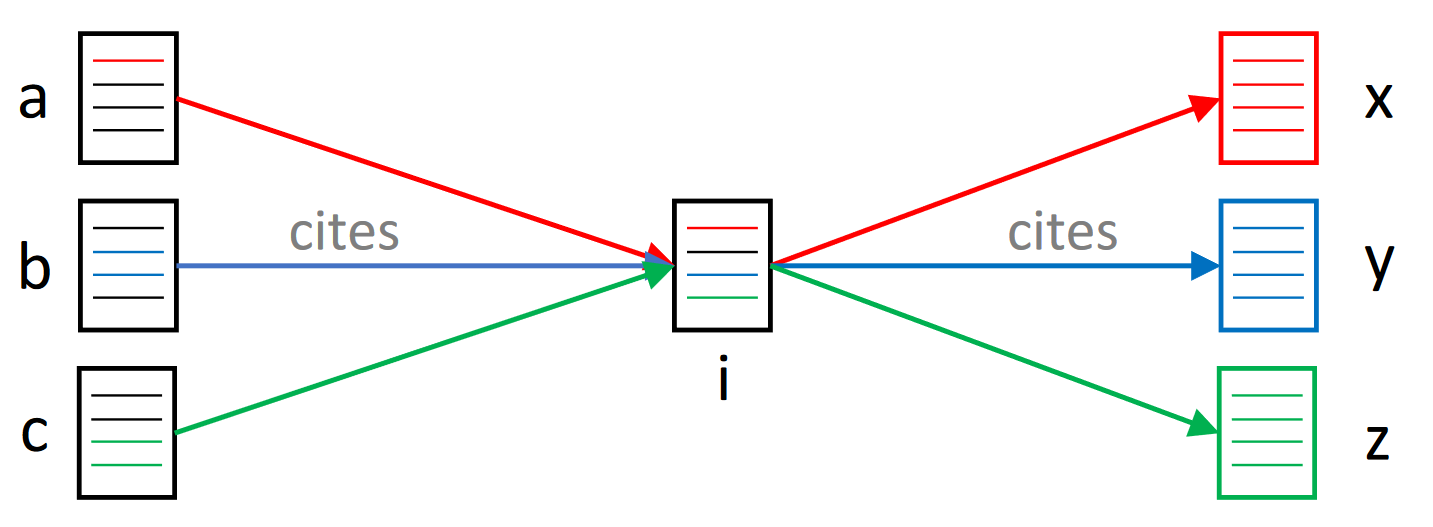}
    \caption{In case of \textit{MAG}, the pseudo full text of paper i consists of its citation contexts for paper x, y, and z.
    In case of \textit{MAG-Cited}, the pseudo full text of paper i consists of citation contexts from papers a, b, and c where paper i is cited.}
    \label{fig:magcited}
\end{figure}

\textbf{MAG with cited contexts (MAG-Cited)}
Huang et al.~\cite{HuangKCMGR12,Huang2015} claimed that ``a citation’s context contains explicit words explaining the citation.'' In other words, the words in a citation context explain the \textit{cited} paper rather than the actual (i.e., citing) paper. We try to verify this claim by creating an alternative data set where each pseudo full text consists of citation contexts taken from papers which cite it, i.e., where it acts as the target of a citation rather than the source (see in Figure~\ref{fig:magcited} the citation contexts in the papers a, b, and c for modeling paper i). 

\textbf{arXiv-MAG}
In contrast to MAG, arXiv and ACL contain the full text of papers. Our evaluation on these data sets aims to determine the effect full text has on citation recommendation.

For arXiv-MAG, we reuse the approach by Saier and Färber~\cite{SaierF19} to convert \LaTeX source files provided by arXiv into text files and create a database in which the papers in the reference section of each paper are mapped to MAG. 
A couple of filter conditions are applied on the data set of arXiv papers: only computer science papers from 1991 to 2017 are considered (until 2016 for training) and papers which have been cited less than five times are discarded. Furthermore, as our final data set will contain additional data from MAG (pseudo full text), it is necessary to map all the arXiv IDs to MAG IDs. All the arXiv papers which cannot be mapped to the MAG are therefore discarded in the training set. 

Pseudo full text is fetched from the MAG for the cited papers which are not within the reduced arXiv data set. 

To prepare the \textbf{test set}, the citation contexts of computer science papers from 2017 are extracted from arXiv. Like in the training set, the arXiv IDs of the cited papers are mapped to MAG IDs, and act as the ground truth. Again, references which are not in the training set are discarded from the ground truth.

\textbf{ACL-MAG}
Färber et al.~\cite{Faerber2018b,Faerber2018} described a method to detect citation contexts from ACL Anthology (and arXiv) papers and provide the corresponding data online. 
They obtained the references of each paper and mapped them to the bibliography database DBLP. 
The data was divided into 3 files for each paper: a text file containing the full text and annotations for citation markers, a references file, and a metadata file. 

To create the \textbf{training set}, we make use of the text files and only consider citation annotations with links to DBLP.  
DBLP papers are mapped to the MAG in the same way as in the case of arXiv. The in-text annotations with DBLP IDs are then replaced by annotations with MAG IDs.
The training data contains all papers from 1965 to 2005. Again, additional MAG pseudo-text is added for these papers' reference papers.

The \textbf{test set} contains citation contexts from ACL Anthology papers in 2006. The papers in the ground truth are fetched from the references and mapped to the MAG in the same way as the training set. Any papers not in the training set are removed from the ground truth. Unlike arXiv, all the test set contexts have only one paper in the ground truth.

\subsubsection{Analysis on Training and Test Sets}
The number of training set papers has interesting ramifications for the evaluation results. In general, it is reasonable to expect that complex algorithms perform better with more training examples. As a result, we hypothesize that less complex algorithms might perform better with fewer training examples. To examine that, the number of training examples varies widely across data sets, with ACL-MAG at the lower end and the regular MAG data set at the higher end. 

Apart from the algorithms themselves, the number of citation contexts in the test sets also have a big effect on the evaluation results. The higher the number of test set contexts, the higher the confidence we can have in the results. The training-test split is done based on years for all the data sets, as shown in Table~\ref{tab:trainingteststats}. The number of test set contexts from these papers varies widely, too, with ACL-MAG being by far the smallest data set in these terms and arXiv-MAG the largest (the ratio of test set contexts to training set papers for arXiv-MAG is also the highest).

\subsection{Evaluation Metrics}

We follow related works and use the following evaluation metrics:

\textbf{Mean Average Precision (MAP)}: The mean average precision is the mean of the average precision values over multiple queries (test set contexts).
Let the average precision of the recommendations for all the $n$ test set contexts be $AP_1, AP_2, AP3, ..., AP_n$. Then, 

\begin{equation*}
    MAP = \frac{\sum\limits_{i=1}^n AP_i}{n}
\end{equation*}

\textbf{Recall@k}:
Recall is a measure which gives the percentage of true positives found in the first $k$ recommendations for a single query. In other words, Recall@k gives the average recall over all $n$ test set contexts:
\begin{equation*}
    Recall = \frac{\sum\limits_{i=1}^n R_i}{n}
\end{equation*}

where $R_i$ is the recall for a single test set context.

\textbf{Mean Reciprocal Rank (MRR)}:
The reciprocal rank is the reciprocal of the rank of a relevant item. The MRR is obtained by averaging the reciprocal ranks over all queries. It is a metric which is most often used when there is only one relevant item. 
As the test sets in this paper \textit{generally} have only one relevant item in the ground truth list, this is a useful metric. 

\textbf{Normalized Discounted Cumulative Gain (NDCG)}:
The discounted cumulative gain assumes that relevant documents have different degrees of relevance. 
Let $rel(r_i)$ be the relevance score. 
Then,
\begin{equation*}
    DCG@k = \frac{rel(r_1)}{1} + \sum\limits_{i=2}^k \frac{rel(r_i)}{log_2 i}
\end{equation*}

where the denominator for i=1 is 1 (as $log_2 1 = 0$). 

The ideal DCG@k (IDCG@k) is calculated by sorting the top $k$ recommendations in descending order and calculating the DCG@k. 

The normalized discounted gain for a single query is obtained by dividing the DCG@k by the ideal DCG@k: 

\begin{equation*}
    NDCG@k = \frac{DCG@k}{IDCG@k}
\end{equation*}

This is divided by the number of queries to get the final mean NDCG@k.

\subsection{Offline Evaluation}
\label{sec:offline-evaluation}

\subsubsection{Evaluation Setting}
We use the following approaches in our offline evaluation:

\begin{itemize}
    \item \textit{Information retrieval methods}: BM25.
    \item \textit{Topic modelling methods}: Variational Bayes LDA used as a baseline (LDA Mallet is included for ACL-MAG) with number of topics as 300 for all the large data sets and 200 for the small ACL-MAG data set.
    \item \textit{Embedding-based methods}: Paper2Vec, Doc2Vec, hd2vOUT (with only OUT document vectors), hd2vINOUT (with IN and OUT document vectors).
    \item \textit{Semi-genetic hybrid algorithms:} \textit{Hybrid} based on BM25 and hd2vOUT, \textit{Hybrid23} based on BM25 and hd2vOUT and using both citing and cited papers (see Section~\ref{sec:hybrid-rec-approach}).
\end{itemize}

Exploratory experiments conducted across data sets indicate a huge variance in performance between approaches. Using all the algorithms as part of the hybrid recommender did not seem to increase recall. Therefore, \textit{Hybrid} and \textit{Hybrid23} combine recommendations from only the BM25 recommender and the Hyperdoc2vec recommender using OUT document vectors (hd2vOUT).

\begin{figure}
    \centering
    \includegraphics[keepaspectratio,width=\linewidth]{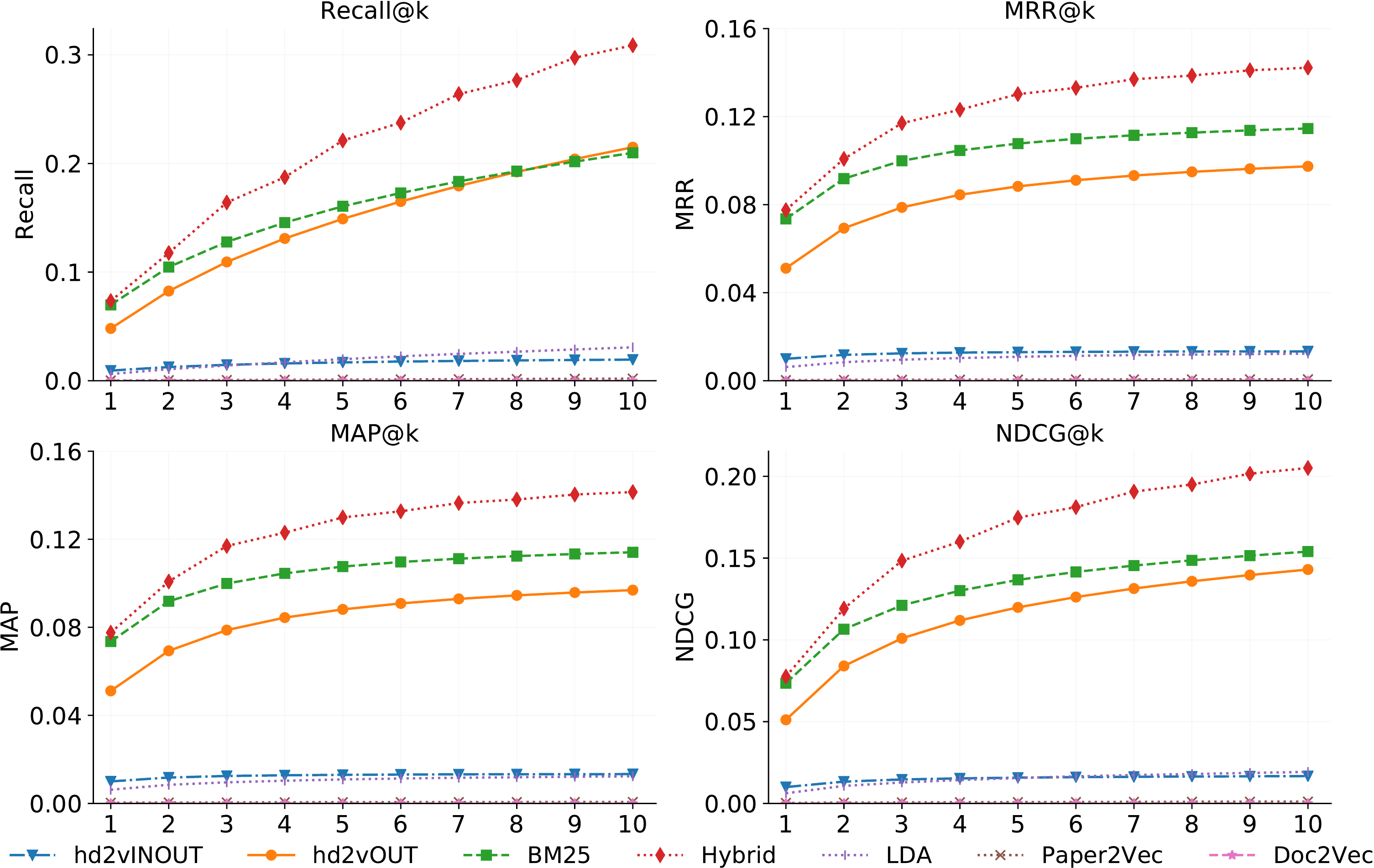} 
    \caption{Evaluation using MAG.}
    \label{fig:magevaluation}
\end{figure}

\begin{table}[tb]
\centering
    \caption{Evaluation scores at $k=10$ for all models using the MAG50 data.}
    \label{tab:mag50evalk10}
\centering
\begin{small}
    \begin{tabular}{lrrrr}
    \toprule
    \textbf{Model} & \textbf{MRR@10} & \textbf{Recall@10} & \textbf{MAP@10} & \textbf{NDCG@10} \\
    \midrule
    BM25       & 0.1528 & 0.2836 & 0.1528 & 0.2082 \\
    LDA     & 0.0369 & 0.0892 & 0.0369 & 0.0567 \\
    Doc2Vec & 0.0000008 & 0.000005 & 0.0000008 & 0.000006 \\
    Paper2Vec        & 0.0025 & 0.0113 & 0.0026 & 0.0055 \\
    hd2vOUT  & 0.1233 & 0.2200 & 0.1233 & 0.1660 \\
    hd2vINOUT & 0.0595 & 0.1353 & \textbf{0.1873} & 0.0898 \\
    Hybrid  & \textbf{0.1873} & \textbf{0.3760} & \textbf{0.1873} & \textbf{0.2711} \\
    \bottomrule
    \end{tabular}
    \end{small}
\end{table}

\subsubsection{Evaluation Results with MAG}

Figure~\ref{fig:magevaluation} shows the evaluation results for all the models tested using the MAG data set without any restriction on the number of citations.

It is clear from the graph that the hybrid model showcases the best results with respect to all used metrics. This is not a surprise as it combines the best parts of the next two best models, BM25 and hd2vOUT. While hd2vOUT has marginally higher recall at $k=10$, BM25 has slightly higher MRR and MAP. 
Unlike on the other data sets, hd2vINOUT performs very badly. This indicates that using the contexts and the abstract as pseudo full text is not enough to generate high quality IN embeddings. The much higher hd2vOUT values show that the OUT embeddings are of much better quality. The OUT embeddings correspond to the papers playing the role of a cited document. Hence, this shows that Hyperdoc2vec is highly context-aware. 

The performance of LDA is below average, but it does better than hd2vINOUT for the MAG data set. Paper2Vec and Doc2Vec perform abysmally across metrics.

\subsubsection{Evaluation Results with MAG50}

One would expect MAG50 to produce better results for all the algorithms on all the metrics, due to the overall smaller candidate pool for each paper. This is indeed the case at $k=10$ (see Table \ref{tab:mag50evalk10}). All the metric values are significantly higher than on all other data sets. This is because the data sets are of similar quality, but the number of candidates for each recommendation is much smaller for MAG50 than for MAG. Consequently, it is more likely that the relevant items are near the top of the list of recommendations.  

However, it cannot be concluded that a system using MAG50 always produces better results. Less popular papers and very new papers are not recommended while using the MAG50 data set even though they may be perfectly valid recommendations. Thus, Hybrid23, our advanced hybrid recommender system, is based on both the MAG and MAG50.

\subsubsection{Evaluation Results with arXiv-MAG}

\begin{figure}[tb]
    \centering
    \includegraphics[keepaspectratio, width=\linewidth]{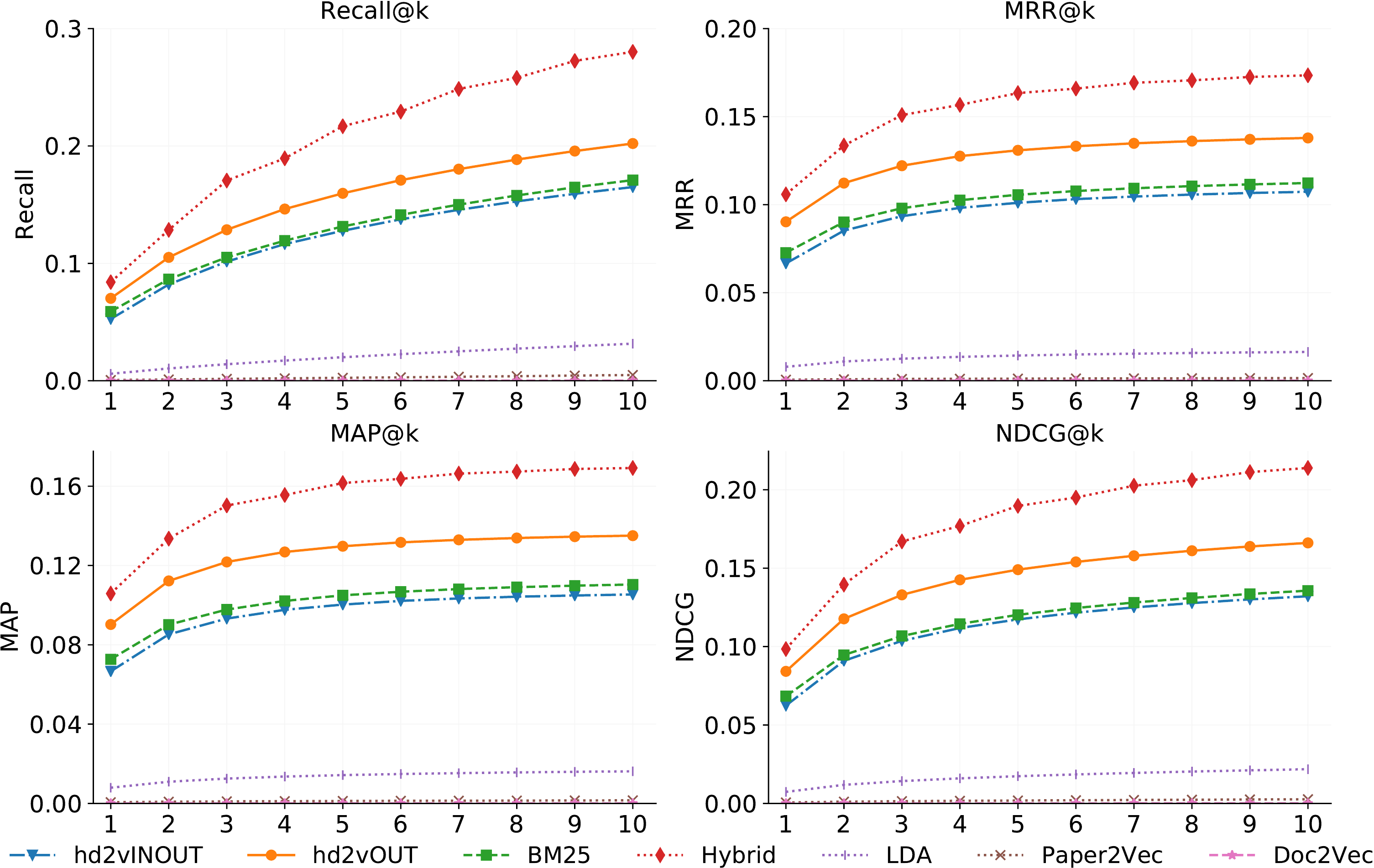}
    \caption{Evaluation using arXiv-MAG.}
    \label{fig:arxivmagevaluation}
\end{figure}

We observe some interesting patterns in the evaluation results for arXiv-MAG from Figure~\ref{fig:arxivmagevaluation}. Unlike when using MAG, hd2vOUT performs better than BM25 across metrics. 
hd2vINOUT performs almost as well as BM25. This proves an important point, namely that Hyperdoc2Vec produces better IN embeddings (IN document vectors) when the full content is available for a large number of papers. The presence of accurate positions of citation markers in the content from arXiv may have played a role in pushing the performance of hd2vOUT over the text-based BM25 algorithm. This is due to the better quality of OUT document vectors. While arXiv-MAG also contains papers with only pseudo full text, it clearly does better with IN embeddings due to the presence of many full text papers from arXiv. 

LDA, again, obtains mediocre results on the metrics. Paper2Vec's and Doc2Vec's performance again show why they were not worth being used in the hybrid algorithm.

Comparing the values of the metrics for arXiv-MAG and the MAG at $k=10$, we see that MAP and MRR for hd2vOUT are higher for arXiv-MAG than for the MAG across algorithms. One important consideration to take into account is that arXiv-MAG has a higher percentage of test contexts with more than 1/2/3 
papers in the ground truth. This improves the probability of getting some of them in the top 10 results, and therefore improves the MAP and the MRR. 

There is not much difference between the performance of BM25 on the MAG and arXiv-MAG data sets in regard to MAP and MRR. This is interesting as it indicates that BM25 performs better on MAG. This could be because a text-based algorithm like BM25 might perform better when the text content is homogeneous across papers. The MAG has pseudo-full text for all papers, while arXiv has full text for some papers and pseudo-full text for others.

The recall is consistently higher across algorithms for MAG than arXiv-MAG. This, again, might have a lot to do with the presence of a lot of ground truths with 4 or more papers in arXiv-MAG. The recall suffers due to this (as opposed to the MAP) as a higher percentage of papers in the ground truth are missing in the top 10 results.

\subsubsection{Evaluation Results with ACL-MAG}

\begin{figure}[tb]
    \centering
    \includegraphics[keepaspectratio, width=\linewidth]{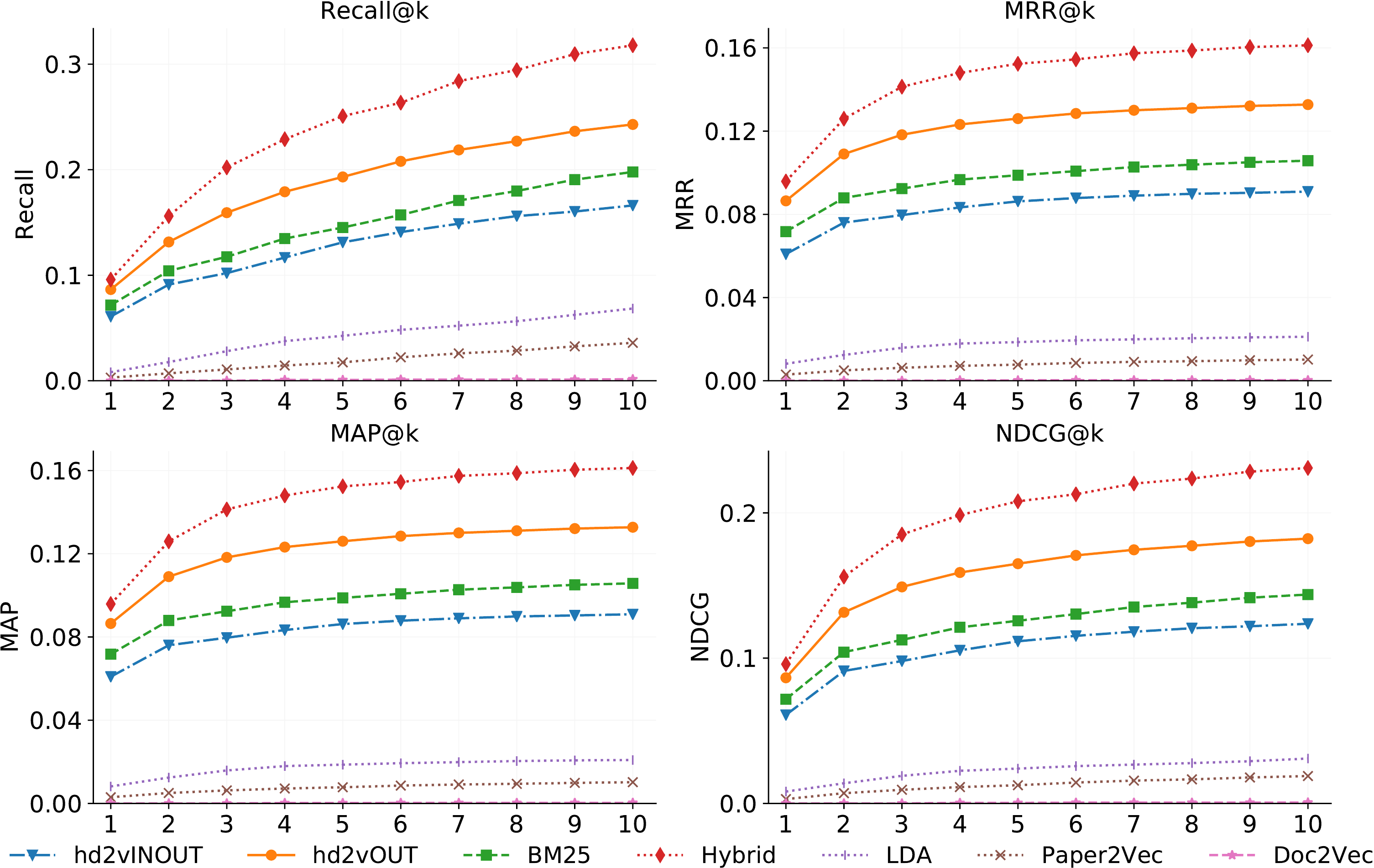}
    \caption{Evaluation using ACL-MAG.}
    \label{fig:aclmagevaluation}
\end{figure}

Moving on to ACL-MAG, which has a very low number of test set contexts, we notice a few common patterns in Figure~\ref{fig:aclmagevaluation}. Method hd2vOUT outperforms BM25 in every metric, which is closely followed by hd2vINOUT. This reinforces the point made that the IN document embeddings are better when some papers have full text. The higher MAP and MRR when compared with the MAG may simply be due to the sheer difference in size of the test sets.

At the end of the offline recommendation phase, we see a clear hierarchy in the performance of the algorithms across data sets and evaluation metrics. The hybrid algorithm performs the best and the two components of the hybrid algorithm, hd2vOUT and BM25 are the second and third best. This is followed by hd2vINOUT and LDA. The other embedding algorithms, Paper2Vec and Doc2Vec, perform very poorly.

\subsubsection{Case Study Using BM25 on 2 MAG Data Sets}
\label{case study}

\begin{figure}[tb]
    \centering
    \includegraphics[keepaspectratio, width=\linewidth]{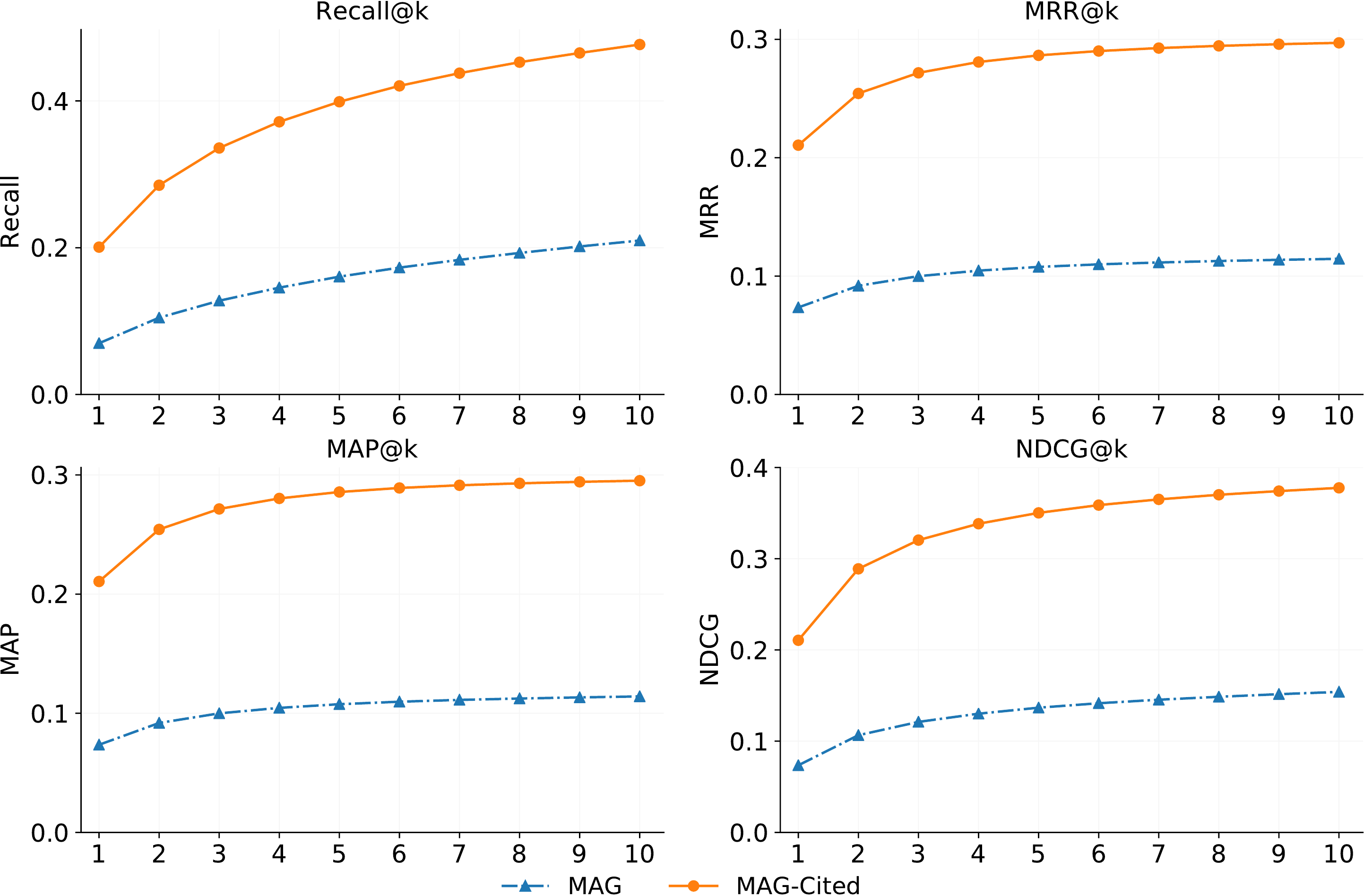}
    \caption{Comparing BM25 performance for the data sets MAG and MAG-Cited.}
    \label{fig:magcitedevaluation}
\end{figure}

In this subsection, we perform a brief case study to test the assertion made in Huang et al.'s two papers~\cite{HuangKCMGR12,Huang2015} that citation contexts describe the cited paper rather than the citing paper. This is done using the BM25 algorithm and the data sets MAG and MAG-Cited.

For the MAG, the pseudo full text for a paper consists of its title, abstract, and citation contexts. 
In contrast, in case of MAG-Cited, a
paper's pseudo full text is made up of its title, abstract, and \textit{citation contexts from papers which cite it}. In other words, the content of a cited paper contains citation contexts from all its citing papers.

Running the BM25 algorithm on both data sets, we find that there is a significant difference in the evaluation results (see Figure~\ref{fig:magcitedevaluation}). The evaluation results show that using the citation contexts of a citing paper as the content of a cited paper pays off. Our evaluation results support Huang et al.'s assertion. 
We incorporate this finding when designing the improved hybrid system \textit{Hybrid23}.

\begin{figure}[tb]
    \centering
    \includegraphics[keepaspectratio, width=\linewidth]{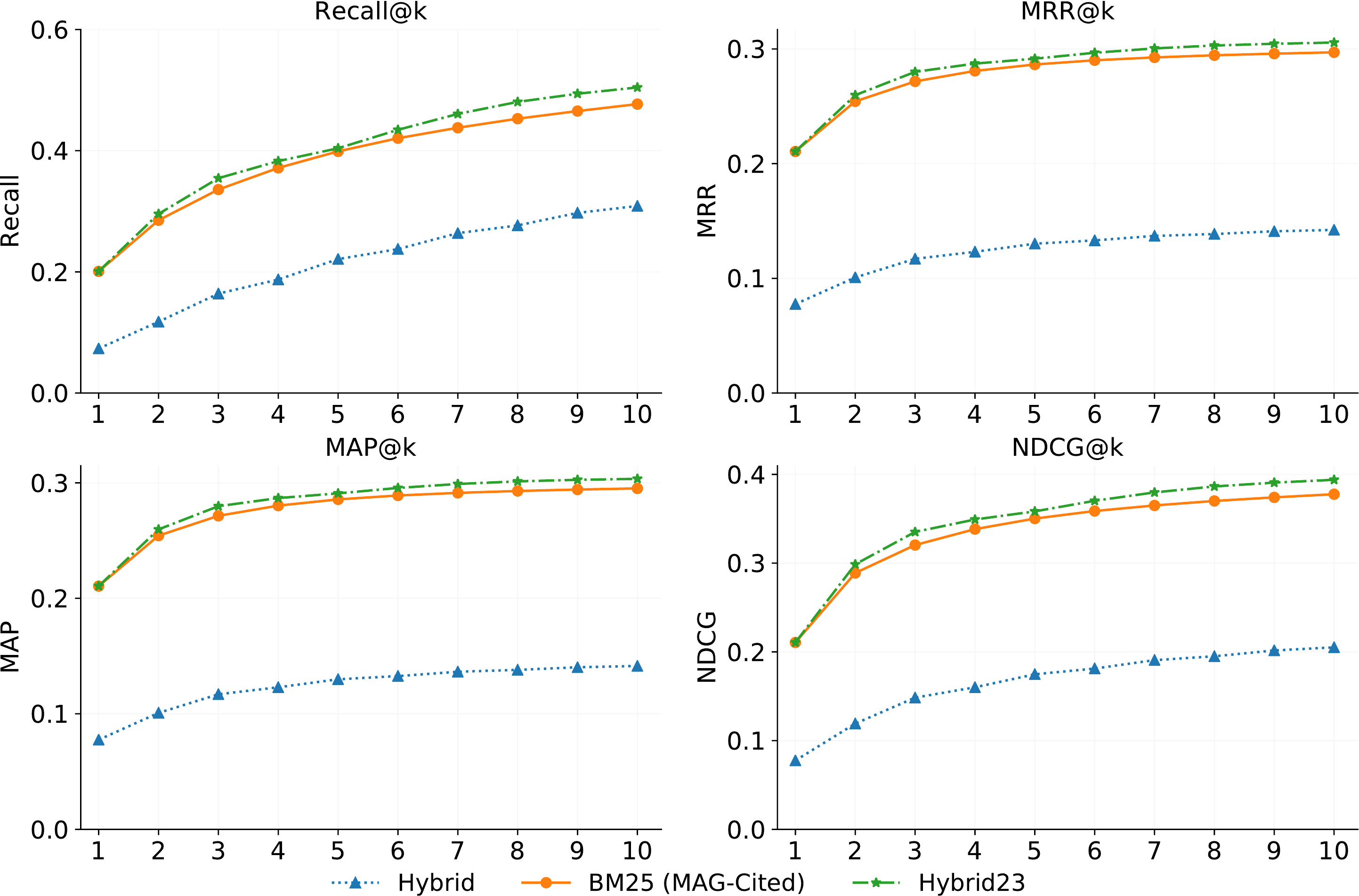}
    \caption{Comparing Hybrid23 with BM25 (MAG-Cited) and the original Hybrid algorithm.}
    \label{fig:maghybridevaluation}
\end{figure}

The graphs in Figure~\ref{fig:maghybridevaluation} compare the original hybrid algorithm with Hybrid23 and BM25 trained on MAG-Cited. Comparing Hybrid23 and BM25 (MAG-Cited), we see that the recall, MAP, and MRR start to get higher for Hybrid23 as $k$ becomes larger. The recall curves, especially, start to diverge around $k=5$. Both algorithms are well ahead of the original hybrid algorithm on all the metrics. 

A quick look at $k=10$ makes it clear that Hybrid23 and BM25 (MAG-Cited) have MAP and MRR values twice as high as the Hybrid algorithm, with substantially higher recall. It is also interesting to note the differences between BM25 (MAG-Cited) and Hybrid23. While there is not too much difference between their MAP and MRR values at $k=10$, a 2.5\% gain in recall is very significant. In fact, the recall@10 is more than 0.5 for Hybrid23, which indicates that a paper from the ground truth is found in the top 10 recommendations in half the test cases.

\subsection{Online Evaluation}
\label{sec:online-evaluation}

We performed a user study in which users (researchers familiar in their research fields) manually rate the recommendations for a number of citation contexts based on the recommendation approaches \textit{BM25}, \textit{hd2vOUT}, and \textit{Hybrid} as well as MAG as data set (with no restriction as to the number of citations). We chose these approaches since they performed best in the offline evaluation on the MAG data set.
We did not include \textit{Hybrid23} in this user study, since we have restricted the user study to the best algorithms which work on a single data set. 

\textbf{Data Set Creation.}
Citation contexts are extracted from the MAG based on language and field of study. English citation contexts are chosen from the natural language processing field as it closely relates to the topic of the paper and the participants' research fields. All the citation contexts are chosen from 2018 and 2019. 
These contexts were seen in the offline evaluation test set. Contexts which cite papers not from the training set are discarded. Duplicate contexts are grouped together. Additionally, citation contexts with 8 or fewer words (stop words not included) are discarded. 

\textbf{Evaluation Process.}
From the 8,356 surviving citation contexts, random sampling is performed to pick 100 citation contexts for the user study. 10 recommendations are made for each of these 100 citation contexts using each of the three chosen algorithms. 
An example can be seen in Table \ref{tab:example-ranking}. The corresponding metadata -- title, abstract, and the published year-- are retrieved from the MAG database. 

\begin{table*}[tb]
    \centering
    \caption{Example of citation recommendation from the online evaluation. The table shows recommendations from our hybrid recommender as well as its component algorithms. 
    Note: \textcolor{blue}{blue} = ground truth; \textcolor{red}{red} = other possibly valid predictions; unhighlighted =  invalid or general recommendations related to the context. It becomes apparent that the actual ground truth papers are included at different ranks in the hybrid and the component algorithms. Other similar papers are recommended as well.} 
    \label{tab:example-ranking}
    \begin{small}
    \begin{tabular}{p{0.55cm}p{5.4cm}p{5.4cm}p{5.4cm}}
    \multicolumn{4}{p{17.7cm}}{\textbf{Citation context:} ``CCG is still able to use a very compact representation of the lexicon, and at the same time to handle non-local dependencies in a simple and effective way. After the release of CCG-annotated datasets [X], there has also been a surge of interest in this formalism within statistical natural language processing, and a wide range of applications including data-driven parsing [Y].''} \\
    \multicolumn{4}{p{17.7cm}}{\textbf{Original citations} [X]: ``CCGbank: A Corpus of CCG Derivations and Dependency Structures Extracted from the Penn Treebank,'' [Y]: ``Wide-Coverage Efficient Statistical Parsing with CCG and Log-Linear Models.''} \\
        \toprule
        \textbf{Rank} & \textbf{hd2vOUT recommendation} & \textbf{BM25 recommendation} & \textbf{Hybrid recommendation} \\
        \midrule
        1 & MaltParser: A Data-Driven Parser-Generator for Dependency Parsing & Chinese CCGbank: extracting CCG derivations from the Penn Chinese Treebank & \textcolor{red}{Wide-coverage efficient statistical parsing with ccg and log-linear models} \\
        \midrule
        2 & \textcolor{red}{Wide-coverage efficient statistical parsing with ccg and log-linear models} & \textcolor{blue}{Rebanking CCGbank for Improved NP Interpretation} & Accurate Unlexicalized Parsing \\
        \midrule
        3 & Discriminative training methods for hidden Markov models: theory and experiments with perceptron algorithms & Shift-Reduce CCG Parsing & Shift-Reduce CCG Parsing \\
        \midrule
        4 & The importance of supertagging for wide-coverage CCG parsing & A New Parsing Algorithm for Combinatory Categorial Grammar & Conditional Random Fields: Probabilistic Models for Segmenting and Labeling Sequence Data \\
        \midrule
        5 & Accurate Unlexicalized Parsing & \textcolor{blue}{Creating a CCGbank and a Wide-Coverage CCG Lexicon for German} & The Proposition Bank: An Annotated Corpus of Semantic Roles \\
        \midrule
        6 & Online Large-Margin Training of Dependency Parsers & \textcolor{blue}{Semi-supervised lexical acquisition for wide-coverage parsing} & \textcolor{red}{CCGbank: A Corpus of CCG Derivations and Dependency Structures Extracted from the Penn Treebank} \\
        \midrule
        7 & Conditional Random Fields: Probabilistic Models for Segmenting and Labeling Sequence Data & A Data-Driven, Factorization Parser for CCG Dependency Structures & \textcolor{blue}{Semi-supervised lexical acquisition for wide-coverage parsing} \\
        \midrule
        8 & The Proposition Bank: An Annotated Corpus of Semantic Roles & A* CCG Parsing with a Supertag-factored Model & A Data-Driven, Factorization Parser for CCG Dependency Structures \\
        \midrule
        9 & Natural Language Processing (Almost) from Scratch & Using CCG categories to improve Hindi dependency parsing & Using CCG categories to improve Hindi dependency parsing \\
        \midrule
        10 & \textcolor{red}{CCGbank: A Corpus of CCG Derivations and Dependency Structures Extracted from the Penn Treebank} & An Incremental Algorithm for Transition-based CCG Parsing & An Incremental Algorithm for Transition-based CCG Parsing \\
        \bottomrule
    \end{tabular}
    \end{small}
\end{table*}

The number of relevant papers for a test citation context is unavailable, unlike in the offline evaluation process. Instead, an approximation is made. The number of relevant results is approximated to be the sum of the results returned by hd2vOUT and BM25. The reason for this approximation is that in a vast majority of cases, the valid results picked by the user for the hybrid algorithm were from the top 10 results of hd2vOUT and BM25. The two individual models often complement each other and pick different results. In many cases, hd2vOUT was found to recommend more general results (e.g. survey papers and often-cited papers), while BM25 was found to recommend more specific results. This approximation only affects the calculation of recall.

We report three metrics for the online evaluation: MAP, MRR, and recall. We do not report NDCG values because the recommendations are binary and not graded. 

\textbf{Evaluation results.} 
The evaluation results are shown in Table~\ref{tab:onlineevalresults}. The proposed hybrid model outperforms the other models by a large margin. They indicate that BM25 plays a dominant role in the hybrid algorithm and outperforms hd2vOUT. However, there are a number of caveats with the online evaluation. The number of test set contexts is very small. The selection of relevant papers by the user is subjective, and the citation contexts are sometimes ambiguous. The recommendations were selected only from one small field of study of computer science -- natural language processing. 

\begin{table}[tb]
    \centering
    \caption{Online evaluation at $k=10$ using the MAG data set.}
    \label{tab:onlineevalresults}
    \begin{small}
    \begin{tabular}{lrrr}
        \toprule
         \textbf{Model} & \textbf{MAP@10} & \textbf{Recall@10} & \textbf{MRR@10}  \\
         \midrule
         BM25 & 0.362 & 0.541 & 0.380  \\
         hd2vOUT & 0.314 & 0.385 & 0.315 \\
         Hybrid & \textbf{0.370} & \textbf{0.680} & \textbf{0.411}  \\
         \bottomrule
    \end{tabular}
    \end{small}
\end{table}

The user study allows us to make a number of inferences. hd2vOUT sometimes recommended general papers relating to the concepts or claims mentioned in the citation contexts. BM25, being a text-based algorithm, recommended more specific results. As a result, their combination in the hybrid algorithm tends to combine the best of both worlds. 

Another observation was that oftentimes, recommendations which are not perfectly suitable for the citation context might still be interesting to the end user. There are some papers which touch on the topic of the citation context, or survey papers about a related research topic that might still be useful for the end user. This \textit{serendipity} is a hallmark of recommender systems in general, and it could help researchers discover new topics related to their research areas. Our hybrid recommender performs very well in this regard, which is not reflected in pure metrics-based analysis alone.

The user study also shows that there are a number of citation contexts for which no recommendations are possible. Finally, it confirms the belief from the offline evaluation that BM25 and such text-based information retrieval algorithms work as well as or better on the citation recommendations task than more complex algorithms.

%% file: 5_conclusion.tex
\section{Conclusion}
\label{sec:conclusion}
In this paper, we adapted and applied several algorithms for local citation recommendation (Paper2Vec, HyperDoc2Vec, LDA, and BM25). 
Furthermore, we combined approaches into the first fully hybrid recommendation approach for local citation recommendation.
For the evaluation, multiple data sets were created and made publicly available. In the evaluation, we demonstrated the superiority of our hybrid approach over the others based on recall, MRR, MAP, and NDCG metrics. 
Based on our findings, we have further improved our hybrid algorithm by incorporating the citing papers' contexts, thereby creating our Hybrid23 algorithm.

For the future, we plan to explore the creation of a machine learning algorithm which can discard more incomplete citation contexts from the pseudo full text and thus improve the quality of the training data. 